\documentclass[a4paper,11pt]{article}
\usepackage{pos}

\title{Exclusive quarkonium photoproduction
in $A$+$A$ UPCs at the LHC in NLO pQCD}

\author*[a,b]{Kari J. Eskola}
\author[a,b,c]{Christopher A. Flett}
\author[a,b]{Vadim Guzey}
\author[a,b]{Topi L\"oyt\"ainen}
\author[a,b]{Hannu Paukkunen}

\affiliation[a]{University of Jyvaskyla, Department of Physics,
P.O. Box 35, FI-40014 University of Jyvaskyla, Finland}

\affiliation[b]{Helsinki Institute of Physics, 
P.O. Box 64, FI-00014 University of Helsinki, Finland}

\affiliation[c]{Universit\'e Paris-Saclay, CNRS, IJCLab, 91405 Orsay, France}

\emailAdd{kari.eskola@jyu.fi}
\emailAdd{christopher.flett@ijclab.in2p3.fr}
\emailAdd{vadim.a.guzey@jyu.fi}
\emailAdd{topi.m.o.loytainen@jyu.fi}
\emailAdd{hannu.paukkunen@jyu.fi}

\abstract{We present the first study of coherent exclusive quarkonium ($J/\psi$, $\Upsilon$) photoproduction in ultraperipheral nucleus-nucleus collisions (UPCs) at the LHC in the framework of collinear factorization and next-to-leading order (NLO) perturbative QCD (pQCD). We make NLO predictions for the $J/\psi$ and $\Upsilon$ rapidity distributions for lead (Pb) and oxygen (O) beams, and quantify their dependence on the factorization/renormalization scale, nuclear parton distribution functions (PDFs) and their uncertainties, and on differences between nuclear PDFs and generalized parton distribution functions (GPDs). We show that within the PDF-originating uncertainties our approach provides a good description of the available $J/\psi$ photoproduction data in Pb+Pb UPCs at the LHC but that the scale uncertainty is significant. We demonstrate that at NLO pQCD the quark contributions are important in the $J/\psi$ case but that gluons clearly dominate the $\Upsilon$ cross sections. We also study how the scale dependence could be tamed by considering O+O/Pb+Pb ratios of the exclusive $J/\psi$ UPC cross sections, and how HERA and p+p/Pb LHC data can help in obtaining better-controlled NLO predictions in the $\Upsilon$ case.}

\FullConference{ HardProbes2023\\
 26-31 March 2023 \\
 Aschaffenburg, Germany\\}

\begin{document}
\maketitle

\section{Motivation for NLO studies}
It was originally proposed by Ryskin \cite{Ryskin:1992ui} that exclusive coherent photoproduction of $J/\psi$ at HERA, $\gamma+\text{p}\rightarrow J/\psi+\text{p}$, probes very efficiently collinearly factorized small-$x$ gluon distributions: The leading-order (LO) cross section is directly proportional to $[xg(x,Q)]^2$ at momentum fractions $x={\cal O}(M^2/W^2)$ and scales $Q={\cal O}(M)$, where $M$ is the the vector meson mass and $W$ is the center-of-mass system energy of the $\gamma$+p system. Our NLO studies of coherent photoproduction of $J/\psi$ \cite{Eskola:2022vpi,Eskola:2022vaf} and $\Upsilon$ \cite{Eskola:2023oos} in $A$+$A$ UPCs at the LHC aim at studying the role of these processes as gluon probes, and at understanding whether one could use these as further constraints in the global analyses of nuclear PDFs. For this, we study the gluon and quark contributions, and uncertainties arising from the scale choices, PDFs and GPD modeling. In particular, it is interesting to see whether the NLO calculation could simultaneously fit the mid- and forward/backward-rapidity measurements better than the previous LO calculations -- see the comparison in Ref.~\cite{ALICE:2021gpt}.   

\section{Theoretical framework}
The theoretical framework here is that of collinear factorization. We compute the rapidity-differential cross sections for coherent photoproduction of heavy vector mesons $V$ in $A_1+A_2$ UPCs as products of the photon fluxes and photon-nucleus cross sections, accounting for the fact that either one of the colliding nuclei can be the photon emitter,
\begin{equation}
        \frac{d\sigma^{A_1A_2\rightarrow A_1VA_2} }{dy} = \left[ k \frac{dN_\gamma^{A_1}}{dk} \sigma^{\gamma A_2 \rightarrow VA_2} \right]_{k=k^+} \\
        + \left[ k \frac{dN_\gamma^{A_2}}{dk} \sigma^{A_1 \gamma\rightarrow A_1 V}  \right]_{k=k^-} 
\label{XS_plus_minus}
\end{equation}
where $k^\pm\approx Me^{\pm y}/2$ are the energies of the photons emitted by the nuclei $A_{1,2}$. The photon fluxes are obtained as impact-parameter integrals of the number of equivalent Weizs\"acker-Williams photons, $N_\gamma^A(k,b)$, at an energy $k$ and impact parameter $\vec b$, multiplied by the Poissonian probability $\Gamma_{AA}(\vec b)$ of having no hadronic interactions, 
\begin{equation}
k \frac{dN_\gamma^{A}}{dk} = \int d^2 b N_\gamma^A(k,b) \Gamma_{AA}(\vec b),
\label{WWflux}
\end{equation}
where $N_\gamma^A(k,b)$ is from Ref.~\cite{Vidovic:1992ik}, and 
$\Gamma_{AA}(\vec b)=\exp[-\sigma_{NN}T_{AA}(b)]$ with $\sigma_{NN}$ being the total nucleon-nucleon cross section and $T_{AA}(b)$ the standard nuclear overlap function computed from the Woods-Saxon nuclear density distribution. Note that there is no cut-off for $b$ in the above integral.
The photon-nucleus cross section for $\gamma+A_2 \rightarrow V+A_2$ (and correspondingly for $\sigma^{A_1\gamma\rightarrow A_1 V}$) is 
\begin{equation}
\sigma^{\gamma A \rightarrow VA} = \frac{d\sigma_{A}^{\gamma N \rightarrow VN}}{dt} \bigg|_{t=0} \int\limits_{t_{\rm min}}^\infty dt' |F_{A}(-t')|^2, \label{eq:xsec1}
\end{equation}
where the nuclear form factor $F_{A}$, originating from the nuclear GPDs, is a Fourier transform of the Woods-Saxon nuclear density distribution, and $t$ is the Mandelstam variable. The per-nucleon cross section, computed here in the $t\rightarrow 0$ limit,  is 
${d\sigma_{A}^{\gamma N \rightarrow VN}}/{dt}= {|\mathcal{M}^{\gamma N \rightarrow VN}_A|^2}/(16\pi W^4)$. Collinear factorization is done here at the amplitude level \cite{Collins:1996fb},
\begin{equation}
\hspace{-0.2cm}\mathcal{M}^{\gamma N \rightarrow VN}_A \hspace{-0.2cm} \propto 
\sqrt{{\langle O_1 \rangle_V}}
\int\limits_{-1}^1 \hspace{-0.1cm}dx [ T_g (x,\xi,\mu_F,\mu_R) F^g_A (x,\xi,t,\mu_F) 
    + T_q (x,\xi,\mu_F,\mu_R) F^{q,S}_A (x,\xi,t,\mu_F) ],
\end{equation}
where $T_g$ and $T_q$ are the pQCD coefficient functions for gluons and quarks, and $F^g_A$ and $F^{q,S}_A$ are the nuclear GPDs (nGPDs), correspondingly ($S$ for flavor singlet). These depend on the momentum fraction $x$,  skewness $\xi$, factorization scale $\mu_F$, and renormalization scale $\mu_R$. The $T_g$ and $T_q$ were computed in NLO pQCD (${\cal O}(\alpha_s^2)$) in Ref.~\cite{Ivanov:2004vd}, and we apply their results here. Note that quarks start to contribute only from the NLO level on. We obtain the non-perturbative NRQCD element $\langle O_1 \rangle_V$ by matching the measured leptonic decay width of the vector meson with the one calculated at NLO. In the forward limit ($t=0$, $\xi=0$), the nGPDs become the usual nuclear PDFs (nPDFs), 
\begin{eqnarray}
F^{g}_A (x,0, 0,\mu_F) &=& F^{g}_A (-x, 0 , 0,\mu_F) = |x|g_A(|x|,\mu_F) \,, \nonumber \\
F^{q,S}_A (x,0,0,\mu_F) &=&  \sum_{q} \left[\theta(x)q_A(x,\mu_F) - \theta(-x)\bar q_A(-x,\mu_F)\right],
\end{eqnarray}
where the sum is over massless quarks, and $\theta(x)$ is the step function. Here, we use the EPPS21 \cite{Eskola:2021nhw}, nNNPDF3.0 \cite{AbdulKhalek:2022fyi} and nCTEQWZSIH \cite{Kusina:2020lyz} nPDFs.

\begin{figure}[t!]
\centering{\hspace{-0.2cm}
\includegraphics[width=0.5\textwidth]{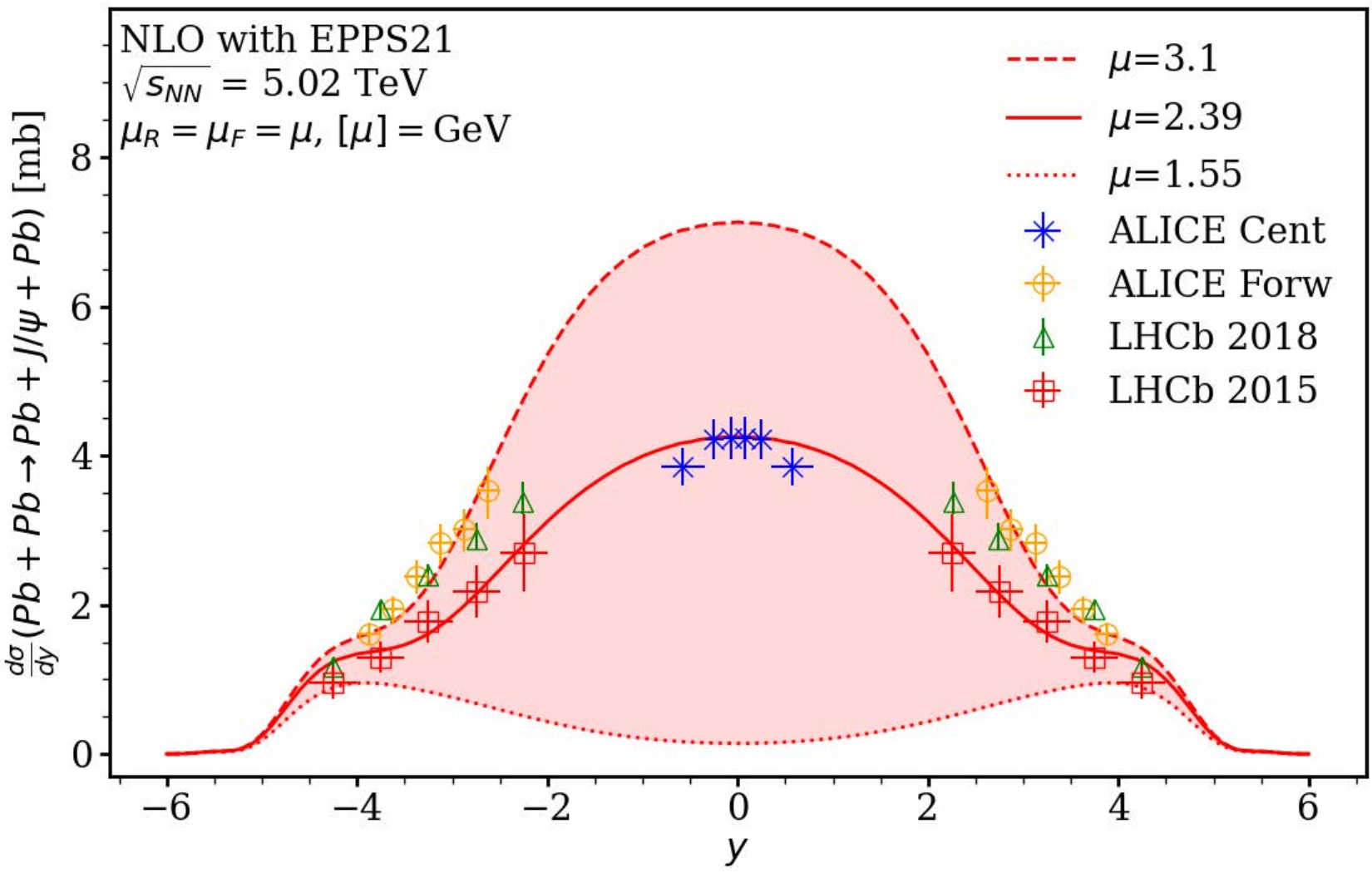}
        \includegraphics[width=0.5\textwidth]{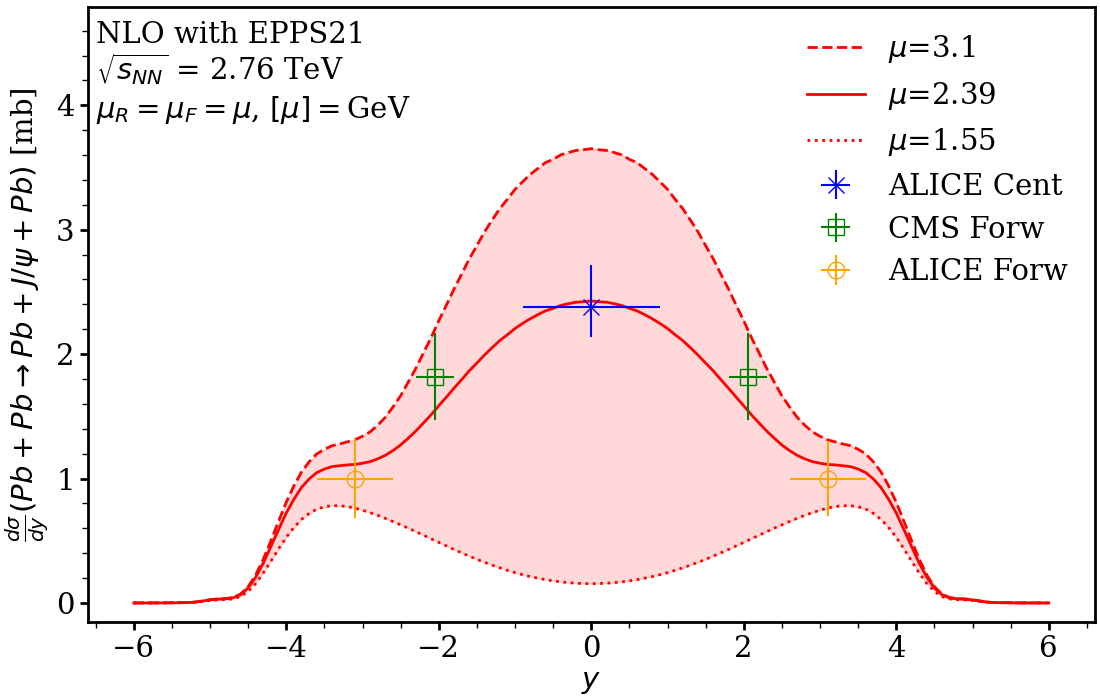}}
\vspace{-0.5cm}
				\caption{\small Rapidity-differential coherent $J/\psi$ photoproduction cross section vs.~rapidity in Pb+Pb UPCs at $\sqrt{s_{NN}}= 5.02$ TeV (\textit{left}) and 2.76 TeV (\textit{right}), obtained with EPPS21 and scales $\mu = M_{J/\psi}/2,\,0.77 M_{J/\psi}$ and $M_{J/\psi}$. For the LHC data references, see \cite{Eskola:2022vaf}. Figure from \cite{Eskola:2022vaf}.}
\vspace{-0.2cm}
\end{figure}

\section{NLO results for $J/\psi$ and $\Upsilon$ in $A$+$A$ UPCs at the LHC}
Taking the nGPDs in the forward limit  and equating the scales $\mu_F=\mu_R=\mu$, we chart the scale dependence of the NLO results by varying $\mu$ from $M_{J/\psi}/2$ to $M_{J/\psi}$. As seen in Fig.~1, the scale uncertainty is significant, a factor of 55 (22) at $y=0$ for the left (right) panel results. With an ``optimal'' scale $\mu\approx 0.77 M_{J/\psi}$ we can, however, fit both the Run 2 and Run 1 LHC ALICE data at $y=0$, although at forward/backward rapidities the NLO calculation has difficulties in reproducing the Run 2 ALICE and latest LHCb data. This, and the HERA data (see comparison in Ref.~\cite{Eskola:2022vpi}) suggests that there is still room for next-to-NLO (NNLO) and NRQCD corrections, as well as effects from GPD modeling.

\begin{figure}[t!]
\vspace{-0.5cm}\hspace{-0.7cm}
\includegraphics[width=.52\textwidth]{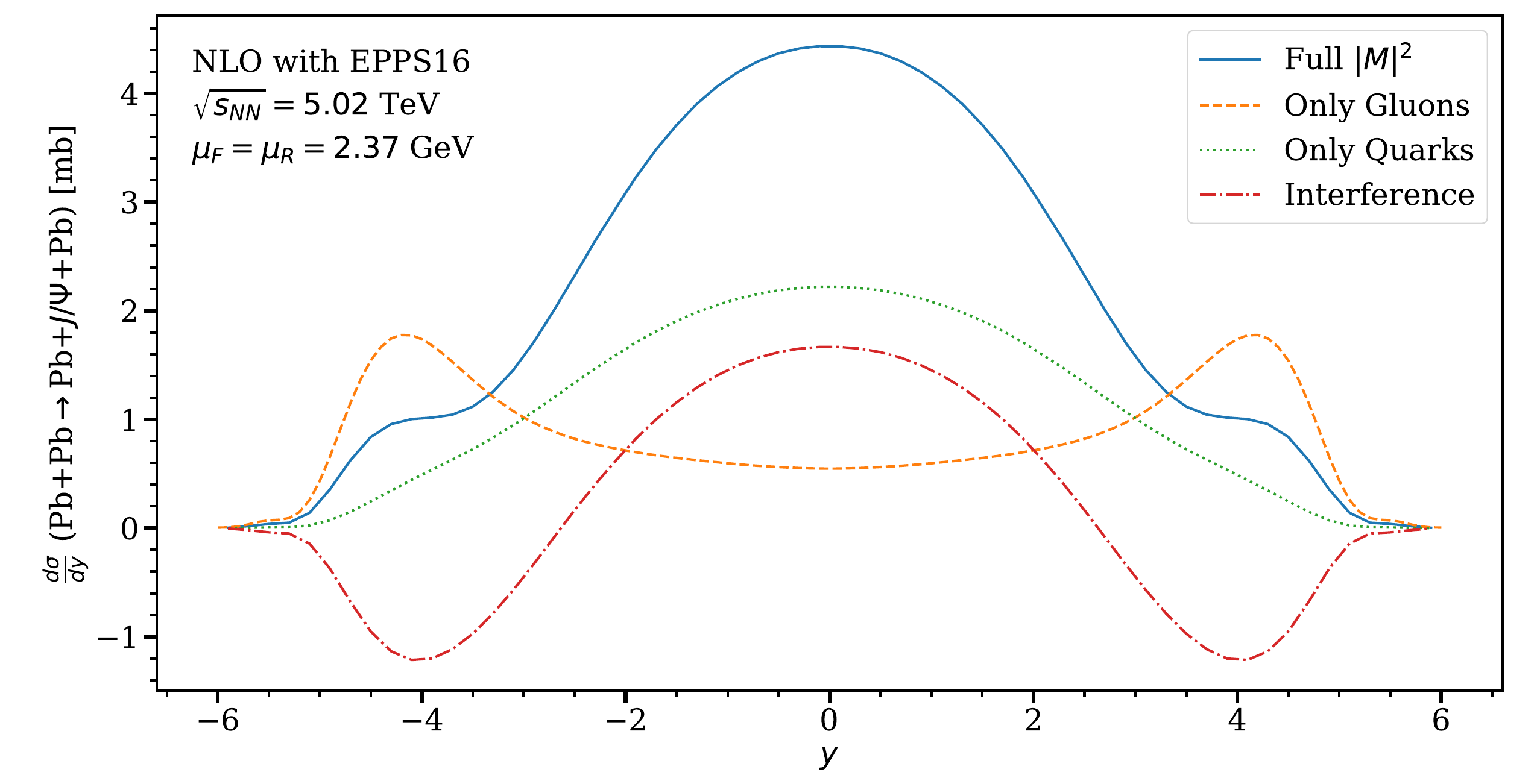}
\includegraphics[width=.54\textwidth]{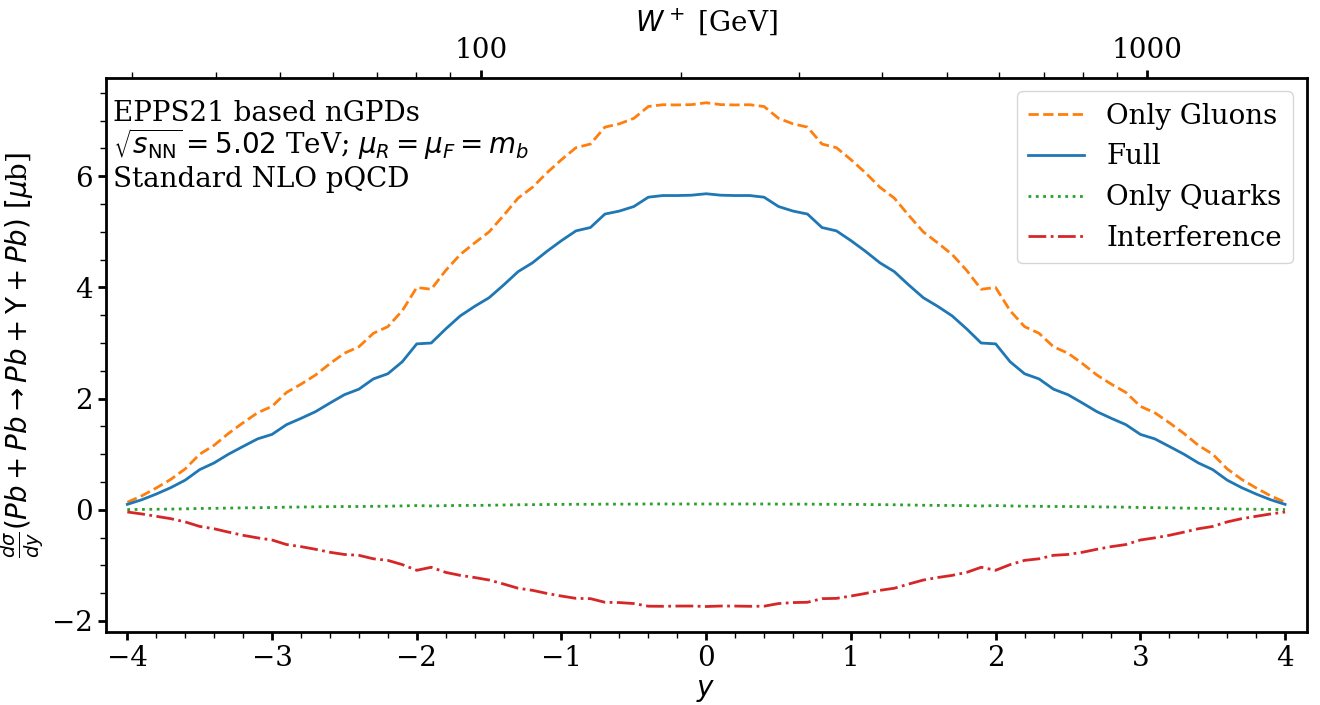}
\vspace{-0.5cm}
\caption{\small Breakdown of the NLO cross section into contributions from quarks only, gluons only, and their interference, for $J/\psi$ (\textit{left}, figure from \cite{Eskola:2022vpi}) and $\Upsilon$ (\textit{right}, figure from \cite{Eskola:2023oos}).}

\vspace{-0.5cm}

\end{figure}

Somewhat surprisingly, see the left panel of Fig.~2, quarks actually dominate over gluons in the NLO cross sections for $J/\psi$ at $y\sim 0$. This is because the LO and NLO gluon amplitudes tend to cancel, which opens room for a significant quark contribution. Because of this the $J/\psi$ cross sections reflect the  gluon nPDFs in a nontrivial way, and certainly they do not scale with the gluon nPDFs squared like in LO \cite{Eskola:2022vpi}. Whether this remains so also at NNLO is yet to be seen. For $\Upsilon$, the gluon dominance is again recovered, as is seen in the right panel of Fig.~2.

\begin{figure}[b!]
\centering{
\includegraphics[width=.8\textwidth]{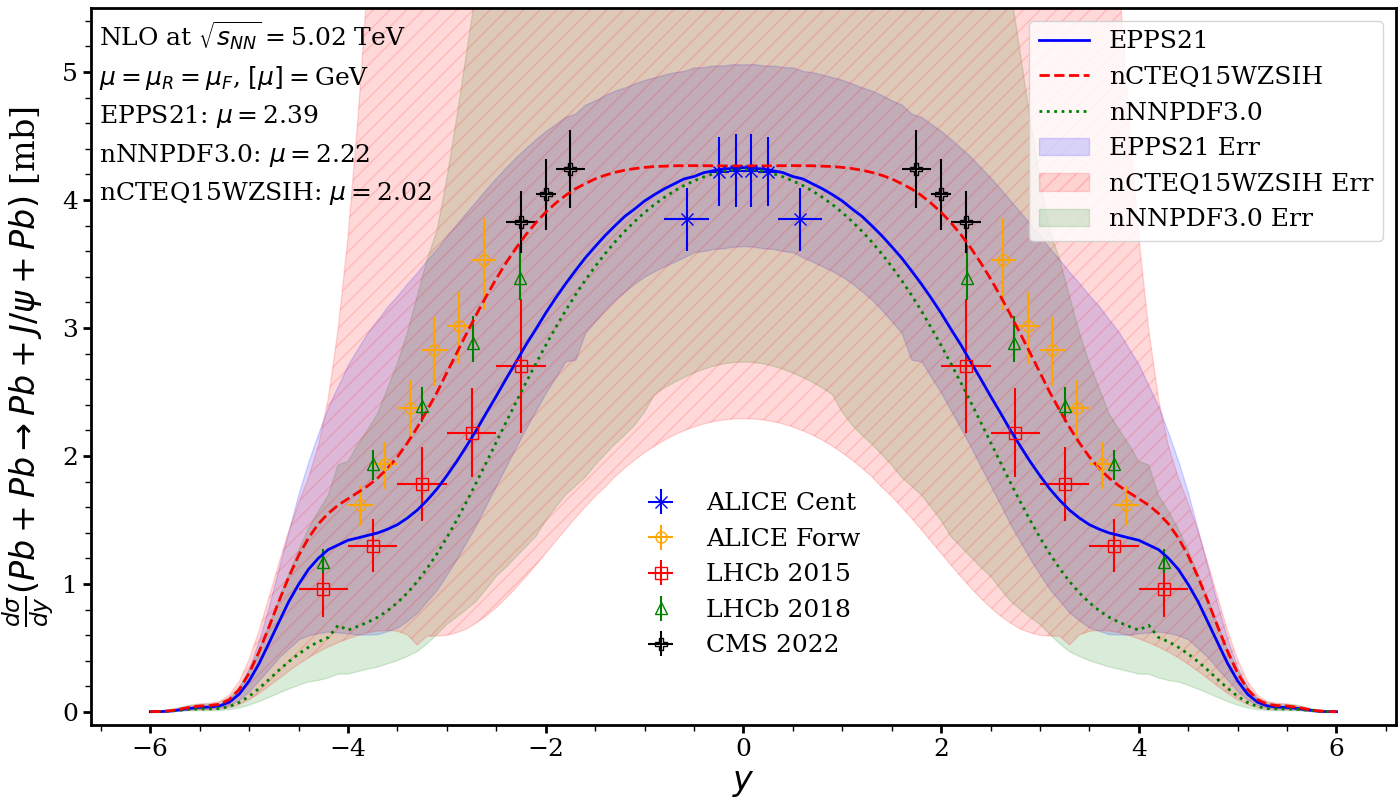}}
\vspace{-0.3cm}
\caption{\small As Fig.~1 left panel, but with PDF uncertainties at the ``optimal'' scale. From \cite{Eskola:2022vaf}, with the new CMS data \cite{CMS:2023snh} added.}

\vspace{-0.5cm}

\end{figure}

The propagation of the nPDF uncertainties to the NLO $J/\psi$ cross section is shown in Fig.~3. The EPPS21 and nNNPDF3.0 error bands include both the nPDF and the PDF uncertainties, while nCTEQ15WZSIH gives only the nuclear ones. The NLO calculation at the optimal scale is consistent with the data within the promisingly moderate EPPS21 error band. The data can be expected to have constraining power on the global fits, provided that the calculation can be brought into a better theoretical control. With the nNNPDF3.0 central set, the NLO $y$-distributions are narrower than those with EPPS21, increasing the tension with the forward/backward data, but interestingly with the nCTEQ15WZSIH central set we do catch also these data. In nCTEQ15WZSIH, the $s$-quark distribution is strongly enhanced, which (with the quark dominance) enhances the cross sections. Thus, the studied $J/\psi$ process may a bit surprisingly turn out to be a probe of the poorly known $s$-quark distributions. Again, whether this feature persists also at NNLO, remains to be seen.

\begin{figure}[t!]
\vspace{-0.2cm}
\hspace{-0.3cm}\includegraphics[width=.55\textwidth]{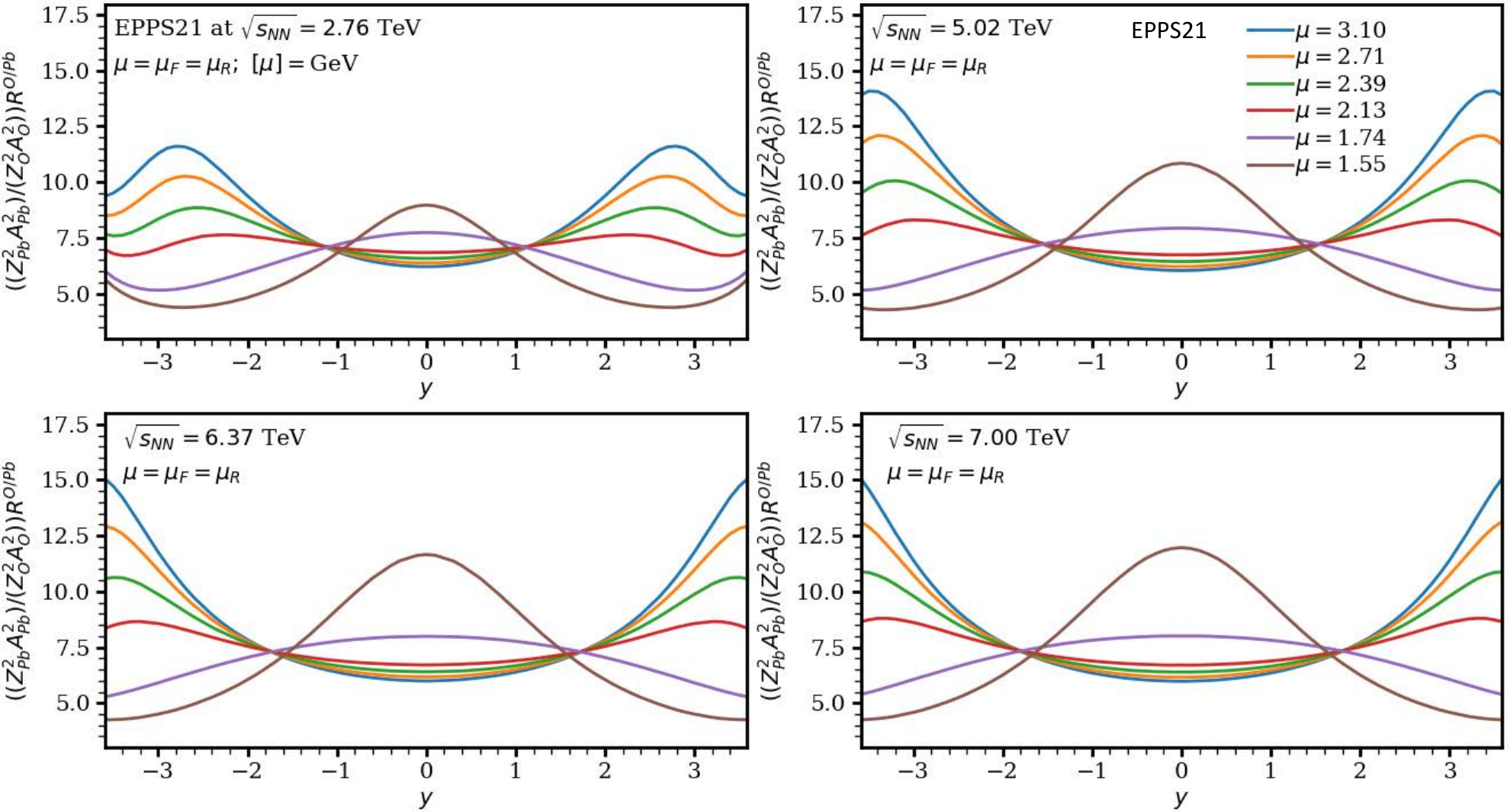}

\vspace{0.cm}\hspace{-0.3cm}
\includegraphics[width=.55\textwidth]{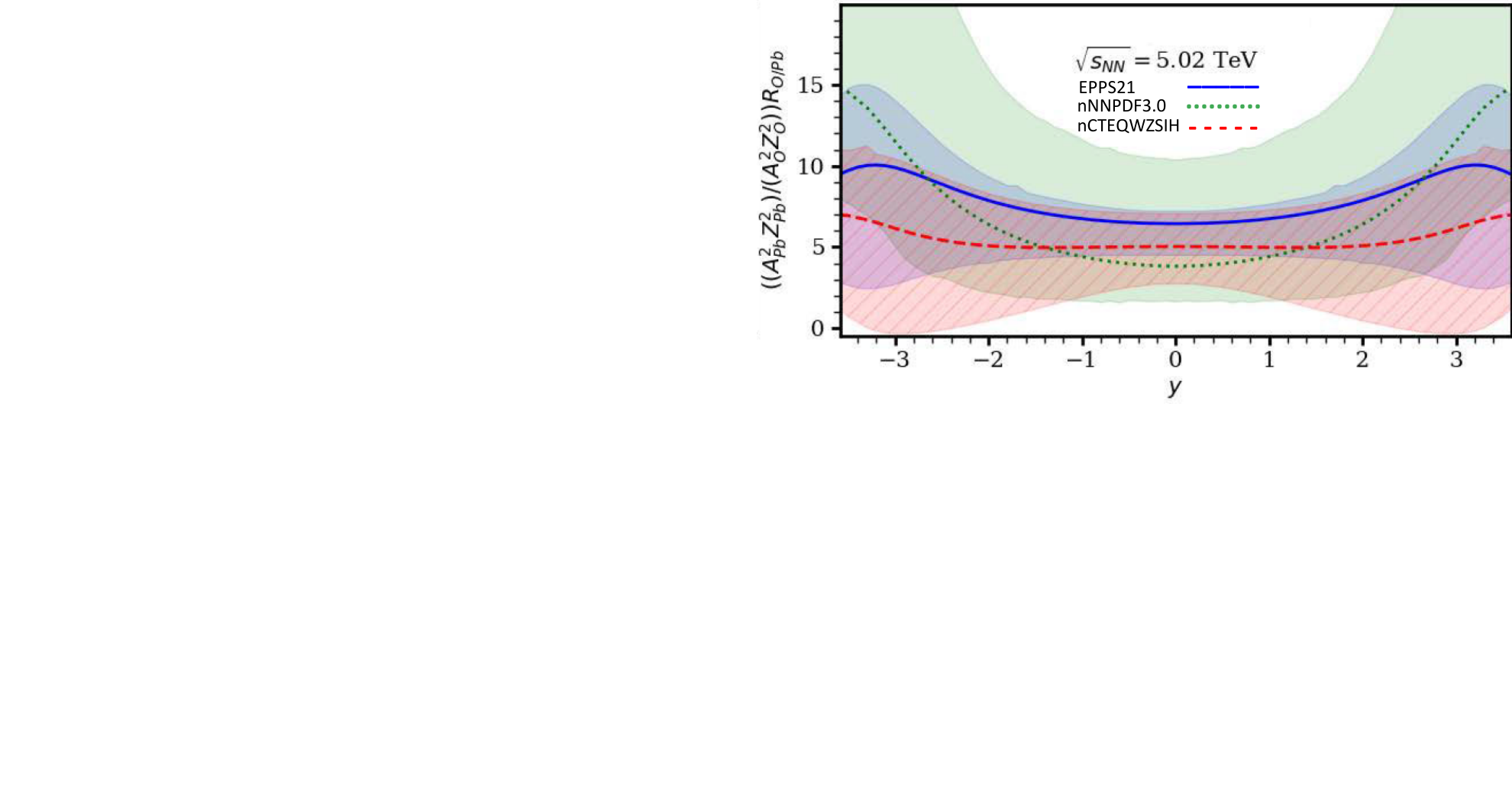}

\vspace{-8.7cm}\hspace{8.3cm}
\includegraphics[width=.5\textwidth]{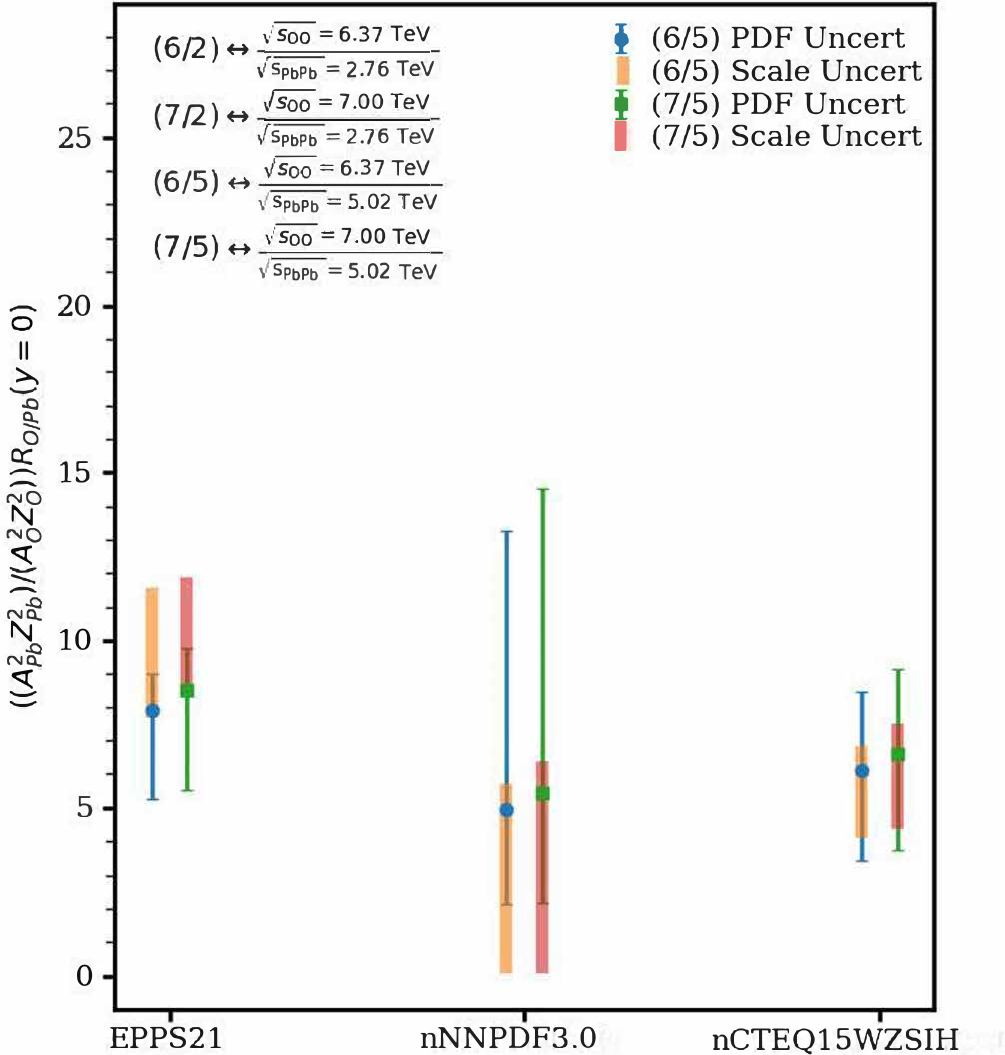}

\vspace{0.5cm}
\caption{\small \textit{Upper left:} Scale dependence of the same-energy ratios of the coherent photoproduction cross sections of $J/\psi$ in O+O and Pb+Pb collisons at the LHC, obtained with EPPS21 nPDFs.
\textit{Lower left:} The same but at the optimal scales of EPPS21, nNNPDF3.0 and nCTEQ15WZSIH, along with their uncertainties. 
\textit{Right:} Different-energy ratios with their scale- and PDF uncertainties at $y=0$.
Figures based on \cite{Eskola:2022vaf}.}

\vspace{-0.5cm}

\end{figure}

In Fig.~4 , we demonstrate that the significant scale uncertainty can be tamed by forming  ratios of $J/\psi$ cross sections in O+O and Pb+Pb UPCs \cite{Eskola:2022vaf}. In the same-energy ratios plotted in the upper left panel, the factor of 55 scale uncertainty at $y=0$ of Fig.~1 shrinks down to a factor of 1.7. The lower left panel shows how these ratios still remain promisingly sensitive to the nPDF uncertainties. Also in the different-energy ratios at $y=0$, shown in the right panel of Fig.~4, the scale uncertainties are at most of the same magnitude as the PDF uncertainties.

Next, we repeat the NLO calculation for coherent $\Upsilon$ photoproduction in Pb+Pb UPCs at the LHC \cite{Eskola:2023oos}. We now also model the GPDs via the Shuvaev transform \cite{Shuvaev:1999fm}. From the $\gamma$+p baseline case shown in the left panel of Fig.~5, we see that HERA $\gamma$+p, and LHC p+p/Pb data favor smaller scales, thus we vary the scale $\mu$ between $M_\Upsilon/4$ and $M_\Upsilon$. As a larger scale process, the scale dependence for $\Upsilon$ becomes clearly weaker than for $J/\psi$. And, as we saw in Fig.~2, gluons now dominate, thus coherent $\Upsilon$ photoproduction in Pb+Pb UPCs probes gluons more directly (at NLO) than $J/\psi$. The left panel of Fig.~5 also shows that the  effects due to the GPD modeling here remain quite small. 
We can also see that the NLO cross sections fall consistently below the HERA/LHC data, indicating again the need for, e.g., NNLO and NRQCD corrections. To make use of the baseline data, we adopt a data-driven method for obtaining the photon-nucleus cross sections, writing
\begin{equation}
\sigma^{\gamma A \rightarrow \Upsilon A}(W) = 
\left[{\sigma^{\gamma A \rightarrow \Upsilon A}(W)}/{\sigma^{\gamma p \rightarrow \Upsilon p}(W)}\right]_\text{pQCD}\sigma^{\gamma p \rightarrow \Upsilon p}_{\text{fit}}(W),
\end{equation}
where the NLO calculation (the pQCD ratio) gives the nuclear effects and the HERA/LHC data fit the normalization. Folding these with the photon fluxes then gives our data-driven predictions shown in the right panel of Fig.~5. Due to the pQCD ratio, the scale uncertainties become now smaller than those of the nuclear GPDs/PDFs, and the GPD effects are seen to become negligible.

\begin{figure}[t!]
\vspace{-0.2cm}
\hspace{-0.7cm}
\includegraphics[width=.5\textwidth]{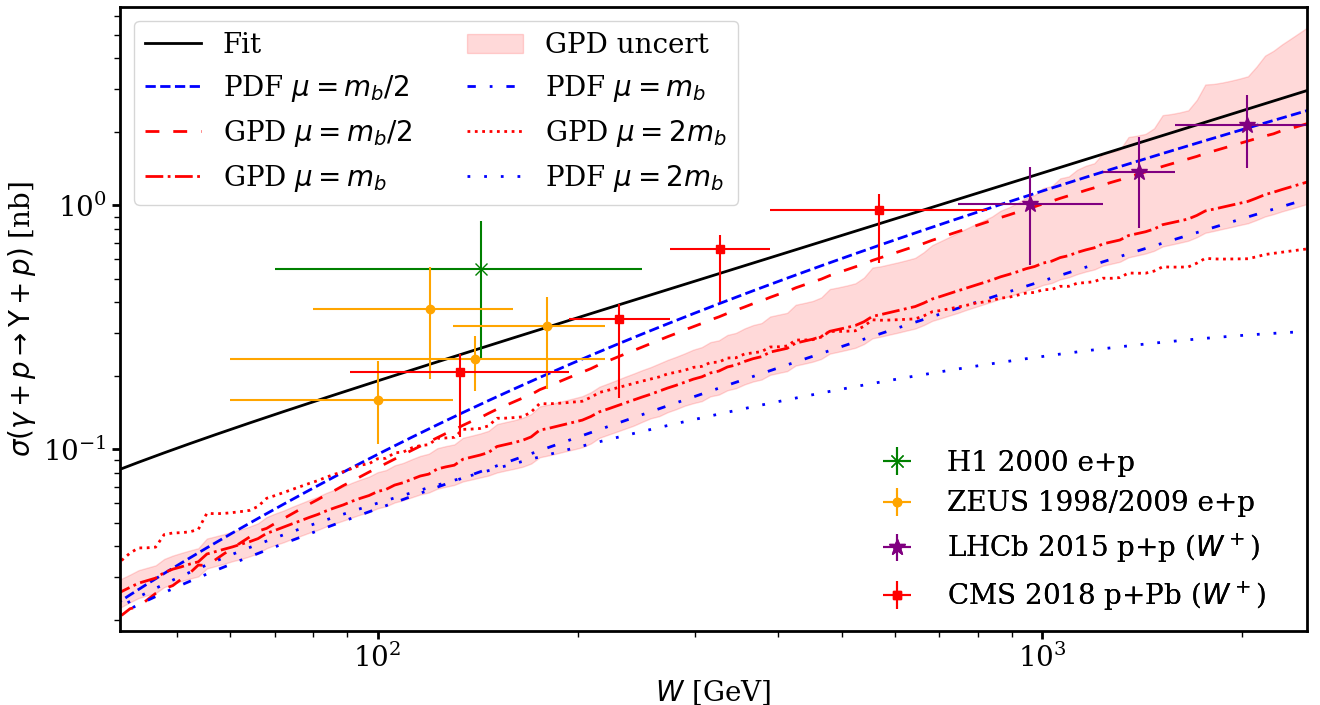}

\vspace{-4.55cm}\hspace{7.0cm}
\includegraphics[width=.57\textwidth]{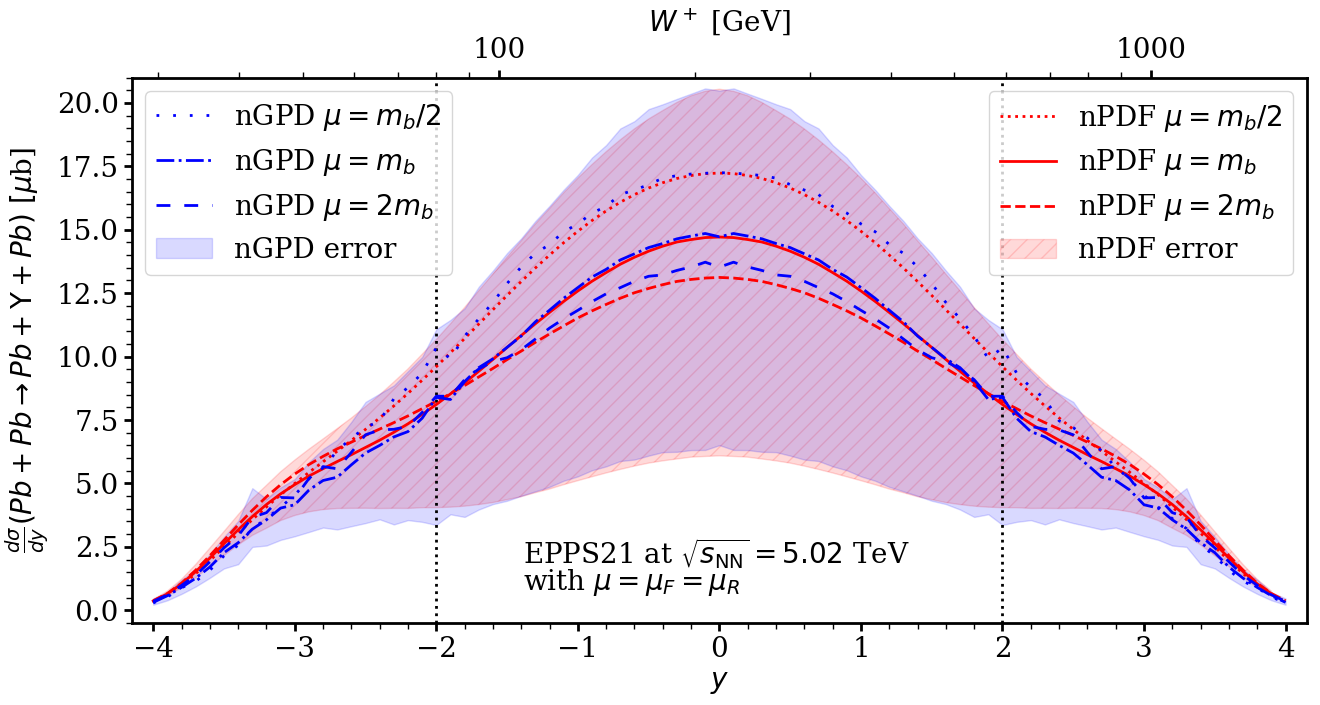}

\vspace{-0.2cm}

\caption{\small \textit{Left:} Coherent photoproduction cross section of $\Upsilon$ in $\gamma$+p collisions, obtained with CT18ANLO-based nGPDs, CT18ANLO nPDFs \cite{Hou:2019efy} and their uncertainties, compared with HERA and LHC data (for data refs., see \cite{Eskola:2023oos}).  
\textit{Right:} Data-driven NLO prediction for coherent photoproduction of $\Upsilon$ in Pb+Pb UPCs at the LHC, for EPPS21-based nGPDs and nPDFs at three scales. The error bands in both panels are for $\mu=m_b$. Figures from \cite{Eskola:2023oos}.}  

\vspace{-0.cm}

\end{figure}

\section{Conclusions and outlook}
We presented the first NLO study of coherent photoproduction of $J/\psi$ and $\Upsilon$ in $A$+$A$ UPCs at the LHC using collinear factorization. For $J/\psi$, a significant scale uncertainty but more moderate nPDF-originating uncertainties were found. Within the nPDF uncertainties, the NLO calculation at a specific optimal scale agrees with all the available LHC data. The PDF uncertainties exceed the data error bars, suggesting PDF-constraining power for the data. For $\Upsilon$, the scale uncertainties are considerably smaller. In both cases, however, there clearly is room for further corrections, such as those arising from NNLO pQCD and NRQCD. The GPD modeling effects via the Shuvaev transform were found to be small in the NLO cross sections. Quarks were found to dominate the NLO cross sections for $J/\psi$, suggesting that $J/\psi$ could be a probe of the elusive $s$-quark distributions. For $\Upsilon$, the gluons were the dominant component, suggesting $\Upsilon$ to be a more direct gluon probe than $J/\psi$. Finally, we made a HERA/LHC-data-driven prediction for the $y$-differential coherent photoproduction cross section of $\Upsilon$, to be tested in the future LHC measurements.\hspace{-0.1cm}\footnote{
We acknowledge the financial support from the MagnusEhrnrooth foundation (T.L.), the Academy  of Finland Projects No. 308301 (H.P.) and No. 330448 (K.J.E.). This research was funded as a part of the Center of Excellence in Quark Matter of the Academy of Finland
(Projects No. 346325 and No. 346326). This research is part of the European Research Council Project No. ERC-2018-ADG-835105 YoctoLHC.}

\end{document}